\newcommand{\grad}{ {\bf \nabla } }

\newcommand{\EQ}{\begin{equation}}
\newcommand{\EN}{\end{equation}}
\newcommand{\EQA}{\begin{eqnarray}}
\newcommand{\ENA}{\end{eqnarray}}

\newcommand{\bm}{\boldmath}

\newcommand{\JJ}{\mbox{\boldmath $j$} {}}
\newcommand{\uu}{\mbox{\boldmath $u$} {}}

\newcommand{\De}      {\mathrm{D}}

\newcommand{\Div} {{{\bm \nabla}}{\bm \cdot}}

\newcommand{\Strain}{\mbox{\boldmath ${\sf S}$} {}}

\newcommand{\Av}      {\mbox{\boldmath $A$} {}}
\newcommand{\Bv}      {\mbox{\boldmath $B$} {}}

\documentclass[twocolumn,twocolappendix]{aastex631}
\usepackage{bm}
\usepackage{amsmath}
\usepackage{booktabs}
\usepackage{hyperref}
\usepackage{mathtools, nccmath}
\usepackage[b]{esvect}

\usepackage{color}  
\usepackage{soul}       
\usepackage[normalem]{ulem}
\usepackage{multirow}

\begin{document}

\title{The Relation between Solar Spicules and Magnetohydrodynamic Shocks}
\correspondingauthor{Sankalp Srivastava}
\email{sankalp.srivastava@iiap.res.in}

\author[0009-0000-2614-254X]{Sankalp Srivastava}
\affiliation{Indian Institute of Astrophysics, $2^{nd}$ Block, Koramangala, Bengaluru-560034, India}
\affiliation{Pondicherry University,
 R.V.Nagar, Kalapet,
 Puducherry-605014, India}

\author[0000-0002-0181-2495]{Piyali Chatterjee}
\affiliation{Indian Institute of Astrophysics, $2^{nd}$ Block, Koramangala, Bengaluru-560034, India}
\affiliation{Pondicherry University,
 R.V.Nagar, Kalapet,
 Puducherry-605014, India}

\author[0000-0002-3369-8471]{Sahel Dey}
\affiliation{School of Information and Physical Sciences, University of Newcastle, University Drive, Callaghan, NSW 2308, Australia}

\author[0000-0003-3439-4127]{Robertus Erd\'elyi}
\affiliation{Solar Physics and Space Plasma Research Centre (SP2RC), School of Mathematical and Physical Sciences, University of Sheffield, Hicks Building, Hounsfield Road, Sheffield, S3 7RH, UK}
\affiliation{Department of Astronomy, Eötvös Loránd University, 1/A Pázmány Péter sétány, H-1117 Budapest, Hungary}
\affiliation{Gyula Bay Zoltán Solar Observatory (GSO), Hungarian Solar Physics Foundation (HSPF), Petőfi tér 3., Gyula, H-5700, Hungary}

\begin{abstract}
Spicules are thin, elongated jet-like features seen in observations of the solar atmosphere, at the interface between the solar photosphere and the corona. These features exhibit highly complex dynamics and are a necessary
connecting link between the cooler, denser solar chromosphere and the extremely hot, tenuous corona. In this
work, we explore the spatial and temporal relation between solar spicules and magneto-hydrodynamic (MHD)
shocks using data from a 2D radiative MHD (rMHD) simulation of the solar atmosphere driven by solar convection. 
Here, we demonstrate, through direct identification, that slow MHD shocks, which propagate along magnetic
field lines, are regions of strong positive vertical acceleration of the plasma that forms the tip of the spicule
material during its rise phase. We quantify the effect of pressure and Lorentz forces on the acceleration of the plasma inside the shocks during the rise of spicules. The causality between spicule and shock propagation in the atmosphere of the model is also investigated. It is further shown that the strength of these shocks may play a vital role in determining the height of the spicules, supporting the idea that shocks act as drivers of some spicules. In addition,
we also find the presence of structures similar to propagating coronal disturbances (PCDs) in
the simulation, linked with the spicules. Here, PCDs appear to be associated with the shock waves driving the spicules that subsequently propagate into the corona and have similar speeds to those reported in observations.
\end{abstract}

\section{Introduction} \label{sec:intro}

Spicules are thin, elongated jet-like features ubiquitously seen shooting upwards in observations of the solar atmosphere, appearing to protrude into the corona before (mostly) falling back to the solar surface \citep{Beckers72,sterling2000solar,DePontieu04,DePontieu2011,tsiropoula2012solar}. These jets show a highly dynamic nature with typical time scales of 2-12 minutes \citep{tsiropoula2012solar}. While spicules were traditionally considered to be chromospheric features, high-resolution observations over the past two decades have established that many spicules show signatures in transition region (TR) and coronal observations as well \citep{DePontieu2011, DePontieu2017ApJL}.

Because of their ubiquitous presence and coupling with multiple layers, they are believed to play a vital role in the mass, energy and momentum transport from the lower atmosphere (photosphere, chromosphere) to the upper layers (TR, corona), and have therefore been suggested as candidates for explaining the heating of Sun's corona to million-Kelvin temperatures \citep{Athay1982ApJ...255..743A,DePontieu2011,samanta2019generation}. Further, it has also been estimated that the entire mass and momentum budget of the solar wind could be accounted for even if only 2\% of the dense spicular material would escape into the heliosphere \citep{withbroe1983ApJ...267..825W}. Thus, understanding these small-scale jet features is key to understanding the solar atmosphere and how its
various layers are coupled. However, partly due to the challenges associated with observing such complex multi-thermal and highly dynamic structures, their physical understanding remains incomplete. 

In parallel to observations, several efforts have been made to study the formation and dynamics of spicules by building theoretical models based on different drivers, e.g. solar global ($p$-mode) oscillations \citep{DePontieu04}, Alfv\'en waves or pulses \citep{oxley2020,Sakaue&Shibata2020ApJ, scalisi2021a, bsingh2022mnras/stac252}, magnetic reconnection \citep{Ding2011A&A}, granular buffeting \citep{roberts1979SoPh...61...23R}, rebound shocks \citep{hollweg1982rebound}, siphon flow due to pressure deficit behind outward shocks from pulse steepening \citep{murawski2010A&A...519A...8M}, and velocity \citep{Kuzma2017A&A}, or pressure pulses \citep{Shibata1982}. It has also been suggested that swirls could excite spicules \citep{liu2019evidence}.

Difficulty with naturally accounting for the heights and abundances of the observed spicules on the basis of few drivers alone has led to the construction of more complex models of spicules that combine several physical mechanisms for excitation. Different mechanisms play a dominant role in the generation of spicules in these models, e.g., \cite{martinez2017generation} proposed magnetic tension release aided by ambipolar diffusion, and \cite{Iijima17} showed the relevance of the Lorentz force in lifting spicule material upwards. 
 
Recently, it has been demonstrated by \cite{dey2022polymeric} that it is possible to account for the observed dynamics of spicules with good accuracy through a model based on solar convection in the presence of vertically imposed magnetic fields, by taking recourse to even simpler physics like radiative transport in local thermal equilibrium (LTE), an ionised equation of state derived from the Saha ionisation equation, and Ohm’s law. 
This model was able to self-consistently excite a forest of spicules with a power law distribution of heights between 6--25\,Mm, in agreement with observations \citep{pereira2014ApJ...792L..15P,skosgrud2015ApJ...806..170S}. Magnetohydrodynamic (MHD) shocks, formed by 
non-linear wave steepening, appeared to be the primary drivers of the simulated spicules in this model. 
This was inferred from the characteristic sawtooth shape of shock waves as seen in acceleration-velocity phase space created by means of Lagrangian tracking. Plasma parcels experience a steep rise in velocity as they enter an acceleration front, and then their velocity drops gradually as the front recedes \citep[see] [Extended Data Fig.~7]{dey2022polymeric}, in line with a sawtooth or N-shaped shock wave profile as a function of time \citep{heggland2007numerical}. 

In this paper, we further examine the data from the 2D radiative MHD (rMHD) simulation reported by \cite{dey2022polymeric} in more detail and investigate the spatio-temporal relationship between 
the simulated solar spicules and associated MHD shocks through direct identification of the shocks. Here, we address the role of shocks in driving the spicules, and quantify the effect of pressure and Lorentz forces on the acceleration of the plasma happening inside the shocks during the rise of spicules. A correlation between simulated spicule heights and strengths of the associated shocks (in terms of acceleration) is also reported. 

Shock waves have previously been explored as drivers of spicule-like features \citep{hansteen2006dynamic,heggland2007numerical,depontieu2007ApJ...655..624D,lrouppe2007ApJ...660L.169R,martinez2009ApJ...701.1569M}. In contrast, in some studies, the jet appears to drive a shock ahead of it. Here, shock wave heating is often conjectured to be produced as a consequence of jetting for explaining enhanced coronal extreme ultraviolet (EUV) emission in connection with spicules e.g., \cite{Klimchuk2012JGRA..11712102K}, \cite{Petralia2014A&A...567A..70P}, \cite{SowMondal2022ApJ...937...71S}. Other examples of previous work that imply 
formation of complex shock fronts due to supersonic jets in the solar atmosphere are \cite{mackenzie2021ApJ...913...19M,mackenzie2022ApJ...929...88M}. In this context we also investigate the causality between spicule and shock propagation in the solar atmosphere using the data from the model reported in \cite{dey2022polymeric}.

A related topic that has been studied is the presence of propagating disturbances in the solar corona, which are upward travelling intensity (density) perturbations that have 
lately been linked with spicules \citep{jiao2015ApJ...809L..17J,samanta2015ApJ...815L..16S,bose2023ApJ...944..171B,Skirvin2024A&A...689A.135S}.  \cite{bryans2016ApJ...829L..18B} 
observationally explored the 
connection between propagating 
coronal disturbances (PCDs) and the 
chromospheric magnetoacoustic shock 
waves that drive jets. \cite{DePontieu2017ApJL} showed that these PCDs might be associated with a mixture of flows (related with spicules) and waves (shock waves 
causing spicule acceleration) at 
lower coronal heights, while at 
upper heights, they might be 
associated mainly with (shock) 
waves. 

Here, we report the presence 
of structures similar to PCDs in the 
given simulation as well, and 
connect them with the presence of 
spicules and shocks. The paper is organized as follows: Section \ref{S-Methodology} describes the numerical 
model and the methods used to 
identify shocks and spicules, while 
the results are presented in Section 
\ref{S-Results}. We conclude with a 
discussion of the results in Section 
\ref{S-Discussion}.

\section{Methodology}
    \label{S-Methodology}
    \subsection{Radiative MHD Simulation Set-up for the Solar Atmosphere}
    \label{S-Simulation setup}

The computational domain (in {\sc Pencil Code}\footnote{\href{https://pencil-code.org/}{https://pencil-code.org/}}) consists of a 2D Cartesian box-in-a-star set-up, covering $-$5 Mm $< z <$ 44 Mm in the vertical direction (with $z=0$ denoting the photosphere), and -9.2\,Mm $< x < 9.2$\,Mm in the horizontal direction. The uppermost part of the solar convection zone is situated in the range -5\,Mm $< z <$ 0, while the atmosphere above spans the photosphere, chromosphere, TR and lower corona. The grid is uniformly spaced, with grid size of 16\,km in each of the two directions, which is sufficient to resolve the approximately hundred kilometers thick transition region of the solar atmosphere.

The 2D simulation data reported in \cite{dey2022polymeric} and further analysed here were obtained by solving the following set of fully compressible rMHD equations on this domain, described in detail in \cite{chatterjee2020testing, dey2022polymeric,Kesri2024ApJ...973...49K}:
    
\begin{equation}
\label{eq:continuity}
\frac{\De\ln \rho}{\De t} \,=\,-{\mathbf{\nabla}\cdot}\uu  + \rho^{-1}(\zeta_D\nabla^2\rho) ,
\end{equation}

\begin{eqnarray}
   \frac{\De\uu}{\De t} \, &\,
   =\,&\,  -\,\frac{\grad p}{\rho}   
      - g_z {\hat{z}}                         
      + \frac{\JJ\times\Bv}{\rho}+\mathbf{F}^\mathrm{corr}_L  
      + \rho^{-1}\mathbf{F}_\mathrm{visc}\, ,
\label{eq:NS}
\end{eqnarray}

\begin{equation}
p\,=\,\frac{\rho R_g T}{\mu(T)}\,,
\label{eq:eos}
\end{equation}

\begin{equation}
\label{eq:induc}
  \frac{\partial\Av}{\partial t}\,
  = \,\uu\times\Bv - \eta\mu_0\JJ +{\bm \nabla}\varPsi\,,
\end{equation}

\begin{eqnarray}
   \rho c_V T\frac{\De \ln T}{\De t}
 \, &\, =\,&\,  -\,(\gamma-1)\rho c_v T \Div \uu+\Div(\mathbf{q}_\mathrm{cond}+\mathbf{q}_\mathrm{rad})
 \nonumber\\
     &\, &\, + \Div(\rho T \chi_t {\bm \nabla} \ln T) + \eta\mu_0\JJ^2
      + 2\rho\nu{\sf{S}}_{ij}^2  \nonumber\\
     &\, &\ +{\mathbf \rho\zeta_\mathrm{shock}\left({\bm \nabla}{\bm \cdot}\uu\right)^2}
      -y_{H}n_{H}^2\varLambda(T).\label{eq:entropy}     
\end{eqnarray}

\begin{equation}
\Div\mathbf{q}_\mathrm{rad}\,=\,\kappa_\mathrm{tot}\rho\oint_{4\pi}(I-S)\,{\mathrm{d}}\varOmega\, .
\end{equation}
Here, $\rho$ is the local plasma density, $\uu$ is the fluid velocity, $\Bv$ is the magnetic field, $\JJ$ is the current density, $p$ is the gas pressure, $\Av$ is the magnetic vector potential, $T$ is the temperature,  $\mathbf{F}_\mathrm{visc}$ is the viscous force, and $\mathbf{F}^\mathrm{corr}_L$ is the semi-relativistic correction to Lorentz force introduced by \cite{boris1970nrl}. 
Also, $g_{z}=2.74 \times 10^{4}$ cm $\text{s}^{-2}$ (constant solar gravity), $R_g=k_B/m_u$ denotes the 
ideal gas constant, $\nu$ 
denotes the kinematic 
viscosity (a function of 
height), and $\Strain$ is 
the traceless rate-of-
strain tensor. An enhanced 
viscous force operates at 
the shock fronts. 
$\zeta_D$ is the 
artificial density 
diffusion coefficient used in the mass continuity 
equation. An artificial density diffusion term is commonly 
employed across a number of 
MHD codes for numerical 
stability e.g., MURaM 
\citep{vogler2005}, Bifrost 
\citep{gudiksen2011}, MANCHA3D \citep{modestov2024}. The effective 
mass, $\mu$, in Eq.~\ref{eq:eos} is determined 
from the number fraction of 
Helium (taking a constant 
value of 0.089), and the 
fraction of ionized 
Hydrogen, $y_{H}$, which is 
calculated using the Saha 
ionization formula 
(assuming LTE), with 
details as given in 
\cite{chatterjee2020testing}. $n_{H}$ denotes the number density of the element Hydrogen.

A vertical magnetic field is imposed in the domain to mimic the effect of large scale polar fields in the coronal hole. This magnetic 
field is built gradually with time, $t$, following 
$B_\mathrm{imp}=B^0_\mathrm{imp}\sin^
2(\pi t/2t_\mathrm{fin})$ over 
$t_\mathrm{fin}=60$\ minutes, 
starting from an initial very small 
value $B^0_\mathrm{imp}=0.354$\,G to reach 74\,G and stay constant thereafter. 
The total resultant 
magnetic field in the 
domain therefore consists 
of the imposed field plus 
the field that develops due 
to the flow, i.e., 
$\Bv=B_\mathrm{imp}\hat{
\mathbf{z}}+\Bv'$ and 
$\Bv'={\boldmath \nabla}\mathbf{\times A}$, where $\eta$ denotes 
the molecular magnetic 
diffusivity. Gauge freedom 
is exploited to set 
$\varPsi=0$ (Weyl gauge) at all times. The reader is 
referred to the Methods section of 
\cite{dey2022polymeric} for more details.

The Spitzer heat conduction flux is represented by ${\mathbf{q}}_\mathrm{cond}$ and described in \cite{chatterjee2020testing}, $\chi_t$ is turbulent diffusion, 
while $c_V$ denotes the specific heat capacity at constant volume. 
The radiative flux is represented as $\mathbf{q}_\mathrm{rad}$ and is calculated using the method of long characteristics as discussed in \cite{heinemann06}. To calculate $\Div\mathbf{q}_\mathrm{rad}$, the radiative transport equation is solved under a grey atmosphere approximation, 
with the source function $S$ given by the 
frequency integrated Planck's function and by neglecting contributions from scattering. $I(x,z,t,\mathbf{\hat{n}})$ represents the specific intensity along the direction of $\mathbf{\hat{n}}$, and the integration is over solid angle, $\varOmega$. 
Power-law fits in analytical form for the Rosseland mean opacity functions are used instead of tabulated opacities. The total opacity, $\kappa_\mathrm{tot}$, is obtained by combining the free-free, bound-free and $\mathrm{H}^{-}$ opacities. For the optically thin plasma in the upper atmosphere, a cooling function is employed, which is available in a tabulated form and calculated using atomic data \citep{Cook89}. 
The initial temperature and density 
stratification is obtained as described in \cite{chatterjee2020testing}. A 
‘sponge’ layer spanning the extent 
32 Mm $< z <$ 44 Mm of the box 
absorbs the outgoing waves without 
allowing them to reflect back into 
the box. The 
temperature at the top boundary, 
$z>32$\,Mm, is held at $10^6$\,K so 
that the atmosphere does not 
collapse while the solar convection 
builds up, since there is no self-
sufficient heating process in our 2D model where Alfvén 
waves are absent and higher 
atmospheric reconnection is 
inhibited by a unipolar coronal 
field. Fixing the temperature at the top boundary is a reasonable approach since we do not wish to address the 
problem of coronal heating directly in this 2D model. A similar approach has been adopted in a number of previous studies \citep{Ijima2015ApJ, rempel2021ApJ...923...79R, Matsumoto2025ApJ}. 

\begin{figure}[ht!] 
\centering
\includegraphics[width=0.4\textwidth]{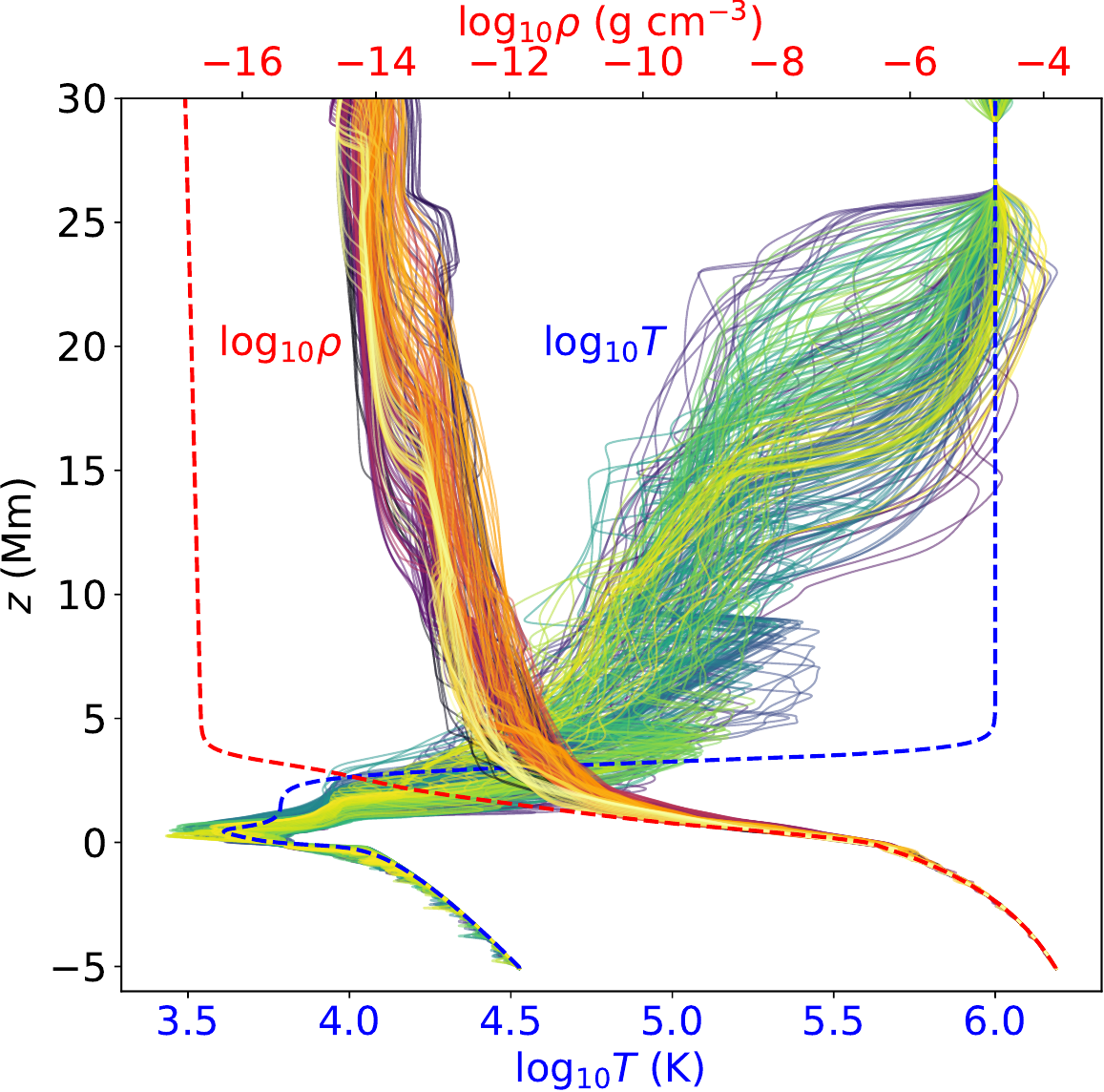} 
\caption{Line plots of density and temperature as a function of $z$ at various fixed $x$ positions from a snapshot at a time $t$=146.7\,min from the start of the simulation. The initial stratification is also shown with the colored dotted lines (red for density, blue for temperature).}
\label{fig:lineplot_rho_temp}
\end{figure}

 The solar convection takes 
nearly 1\,hr of solar time to 
saturate. In this work, we use the 
results corresponding to 20\,min of 
data taken after the convection has 
saturated. We studied 118 snapshots 
with a cadence of 10~s from the 
simulation with a vertically imposed 
magnetic field, $B_{\text{imp}}$, of 
74~G. The first snapshot analyzed corresponds to 
time $t$=146.7\,min from the start of the simulation, and the last one corresponds to $t$= 166.2 min from the start. The 10\,s cadence resolves the lifetime and dynamics of the synthetic spicules ($\sim$ 12-14\,min) very well. Further, this cadence is also suitable for studying the plasma dynamics connected to the MHD shocks being investigated here. This is because the calculation of dynamical parameters like plasma velocity and acceleration carried out here is independent of the cadence, while on the other hand, the quantities that depend on the cadence are the shock propagation speeds as estimated from the slopes (in Fig.~\ref{fig:shock_speed}, for example). Since the slopes are reasonably smooth at 10\,s cadence, any further increase in cadence is unlikely to change the propagation speeds. The initial conditions of the simulation setup along with relaxed values of density and temperature are shown in Fig.~\ref{fig:lineplot_rho_temp}.

\subsection{Visualizing Spicules: Synthetic Emission}
\label{S-Syn-int}

For the visualization of spicules, we utilized the concept of synthetic emission. The synthetic intensity for a line seen in transition region or upper chromosphere can be found from the expression \cite[valid for temperatures between $10^4$--$10^8$\,K, please see][] {Landi08}:
\begin{equation}
\label{eq:syn}
I_\lambda=\int G_\lambda(\rho, T) \varphi(T) dT,
\end{equation}
where $G_{\lambda} $, the ``contribution function" for the line, has been approximated here by the simple analytical expression
\begin{displaymath}G_\lambda(\rho, T)=\exp\left[-\left(\log(T/T_\lambda)/w\right)^2\right] ,\end{displaymath}
with $w=\log 1.8 $. We use $T_{\lambda} = 15,000 \text{K}$ (that corresponds to the Mg II k line, but under the optically thin approximation here). Even if this calculation of emission using an analytical contribution function is not accurate, the dynamics of the modeled spicules will be reasonably reflected.

\begin{figure}[ht!] 
\includegraphics[width=0.5\textwidth]{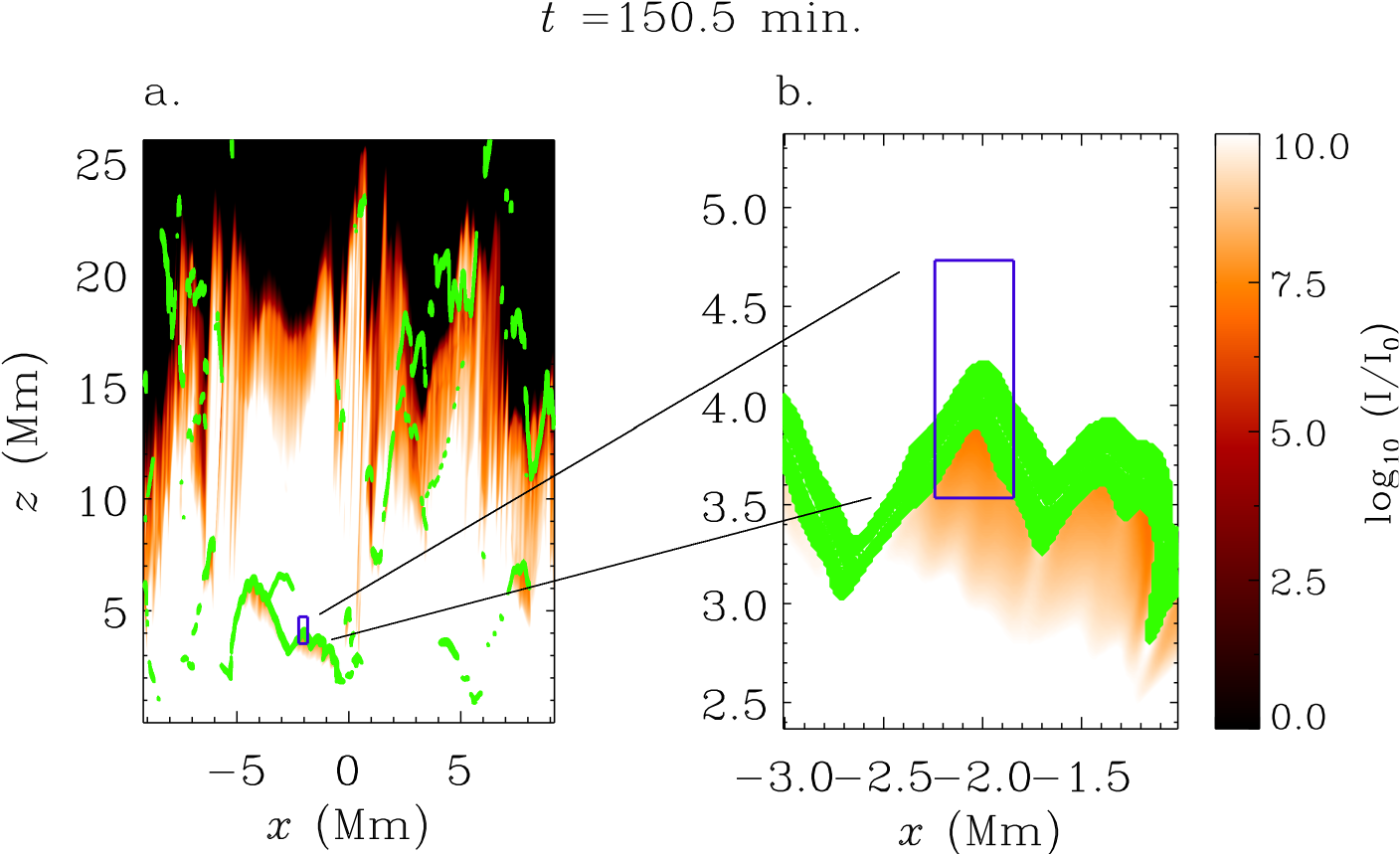} 
\caption{One of the selected cases. Synthetic intensity (for emission at $T_{\lambda} = 15,000 \text{K}$) at a given time from the start of the simulation is shown, as indicated. The shock fronts are overplotted in green, and the region selected for analysis is marked with a box (blue). Panel (b) is an inset of (a), showing the selected region more closely.
}
\label{fig:selected_case_inset}
\end{figure}

The quantity  $\varphi(T)=(\rho/\overline{\rho})^2 ds/dT$ denotes the differential emission measure, scaled here by the square of the horizontally averaged density, $\overline{\rho}(z)$. However, for the 2D simulation run discussed here, the line-of-sight (LOS) integration ($ds$ being an infinitesimal element along LOS) as defined by Eq.~\ref{eq:syn} and the definition of $\varphi(T)$, is not required, for only one grid point exists along the LOS. Therefore, the expression   
\begin{equation} \label{eq:3}
I_{\lambda}=  G_{\lambda} (\rho, T) \Big(\frac{\rho}{\overline{\rho}} \Big)^{2} 
\end{equation}
was used for obtaining the synthetic intensity, which is the result of combining Eq.~\ref{eq:syn} and the definition of $\varphi(T)$, and ignoring the LOS integration for 2D. The synthetic intensity plots for each snapshot were generated after carrying out the calculation (see Fig.~\ref{fig:selected_case_inset}).

\subsection{Identification of Shock Fronts and Calculation of Mach Number}\label{S-shock_ident}
\begin{figure}[ht!] 
\includegraphics[width=0.5\textwidth]{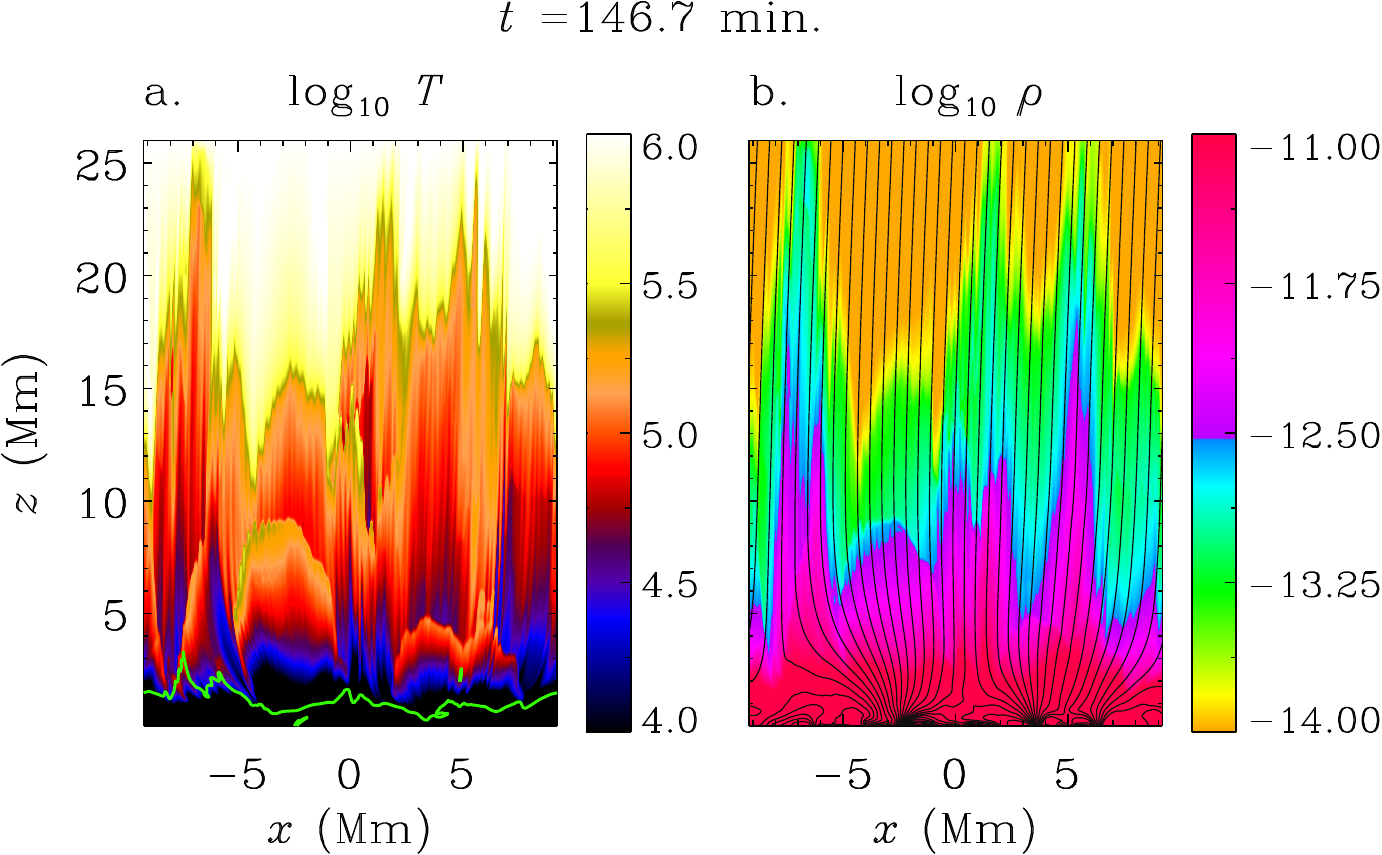} 
\caption{(a) Temperature and (b) density plots at a given time from the start of the simulation, as indicated. The plasma-$\beta=1$ surface is overplotted with the green line in (a), while the magnetic field lines are overplotted in black in (b).  Animation of this figure for the total duration of around 20 minutes (from $t=146.7$~min. to $t=166.2$~min.) is provided in the online journal.
}
\label{fig:beta}
\end{figure}

We detected shock fronts in our simulation snapshots by locating regions with large positive values of flow convergence ${\mathbf{-\nabla}\cdot}\uu $ (and thereby strong local compression of plasma). Shock fronts are identified as regions where ${\mathbf{-\nabla}\cdot}\uu $ exceeds a certain (positive) threshold value. The threshold was fixed as follows. We first calculated, for each snapshot, the quantity $0.4 ({\mathbf{-\nabla}\cdot}\uu)_{\text{max}} $, where $({\mathbf{-\nabla}\cdot}\uu)_{\text{max}} $ denotes the maximum value of ${\mathbf{-\nabla}\cdot}\uu $ within the range 1\,Mm $< z <$ 10\,Mm for that snapshot. Then the minimum of this quantity across all snapshots was chosen as threshold for identifying shock fronts. The shock fronts emerging from below become prominent and well defined between heights 3-5\,Mm. 
The plasma-$\beta$, that denotes the ratio between gas and magnetic pressure, also falls to values below 1 at these heights (see Fig.~\ref{fig:beta}).

Next, we selected certain shock fronts from these snapshots for further analysis (from the lower heights where the fronts become well defined). 
The shock fronts for different snapshots, identified based on our shock detection criterion mentioned above, were visually inspected and few clearly visible cases (14 in total), were picked out. 
The number of selected cases was limited by the total time duration of available data ($\sim$ 20\,min). A typical exemplary case is shown in Fig.~\ref{fig:selected_case_inset}. The region selected for analysis is demarcated by a blue box. The criterion on $-{\mathbf{\nabla}\cdot}\uu $ used for identifying shock fronts takes into account the compression in both $x$ and $z$ directions. In our model, in each of the 14 cases, the compression region thus identified has strong spatio-temporal correlation with the vertical pressure gradient. This rules out any shocked region wrongly identified due to compression in only the $x$-direction.

\begin{figure*}[ht!] 
\includegraphics[width=\textwidth]{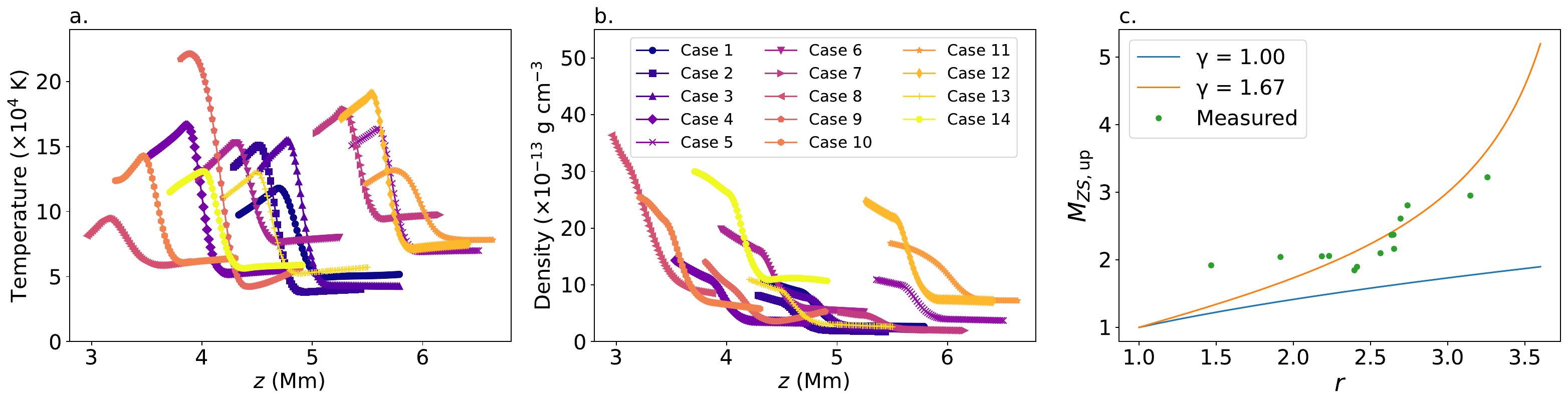} 
\caption{(a) Temperature and (b) density jumps across the shock region, for the 14 selected cases shown in Table \ref{table:shock_slit}. The legend denotes the cases in the table. Panel (c) shows the compression ratios and corresponding upstream Mach numbers from measurements for the 14 cases (green filled circles), along with the theoretically calculated upstream Mach numbers from Eq.~\ref{eq:machno_jump_inverted} for adiabatic ($\gamma=1.67$, orange line) and isothermal ($\gamma=1.0$, blue line) shocks. 
}
\label{fig:jumps_RH}
\end{figure*}

The shock detection criterion used in the study is further validated for the selected cases as follows. In each of the selected cases, we located the tip of the compression (shock) front, and then examined the values of density and temperature, which change as we move across the front (along a vertical line at the horizontal location of the tip). The density and temperature as a function of $z$ are shown in the Fig.~\ref{fig:jumps_RH}. The jumps or discontinuities in the values of these quantities from upstream to downstream region of the selected front in each case can be clearly seen. The density jumps lie in the range 1.5--3.3 (see Table \ref{table:shock_slit} and also section \ref{S-RH cond}).

\begin{figure}[ht!] 
\includegraphics[width=0.48\textwidth]{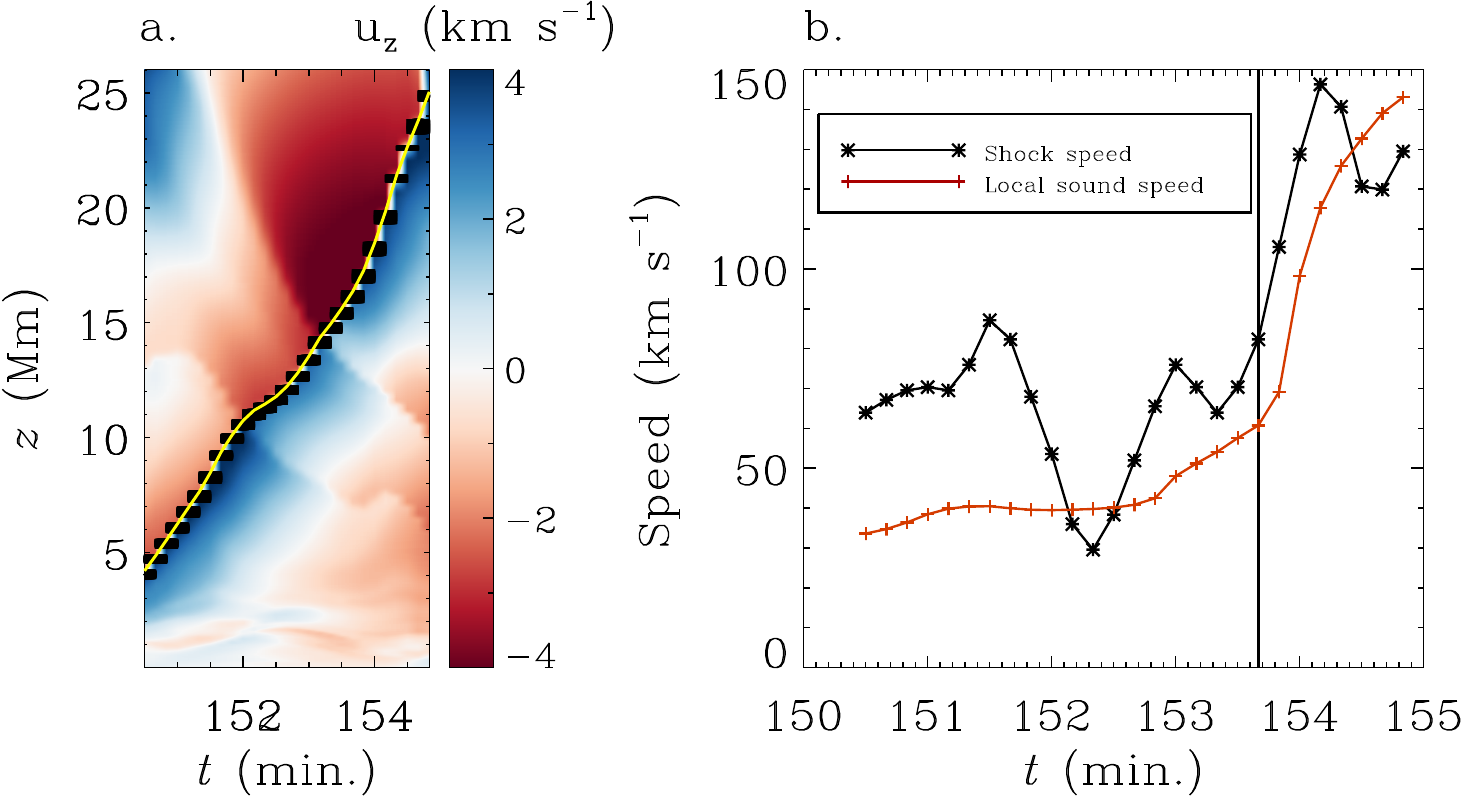} 
\caption{(a) Time-distance plot of the $z$-component of plasma velocity along a vertical slit for a typical selected case (case~4) from Table~\ref{table:shock_slit}. The rectangular black symbols denote the position of the shock front. The finite $z$ extent of the symbols provides the vertical width of the shock fronts along the slit. The $z-t$ curve (in yellow) follows the shock front tip. (b) The shock speed as a function of time (black asterisk symbol) is derived from the yellow $z-t$ curve in panel (a). The vertical line marks the time when the shock front and spicule begin to separate. The sound speed of the local plasma at the shock front tip as a function of time is shown with red plus symbols.}
\label{fig:shock_speed}
\end{figure}

\begin{figure}[ht!] 
\includegraphics[width=0.5\textwidth]{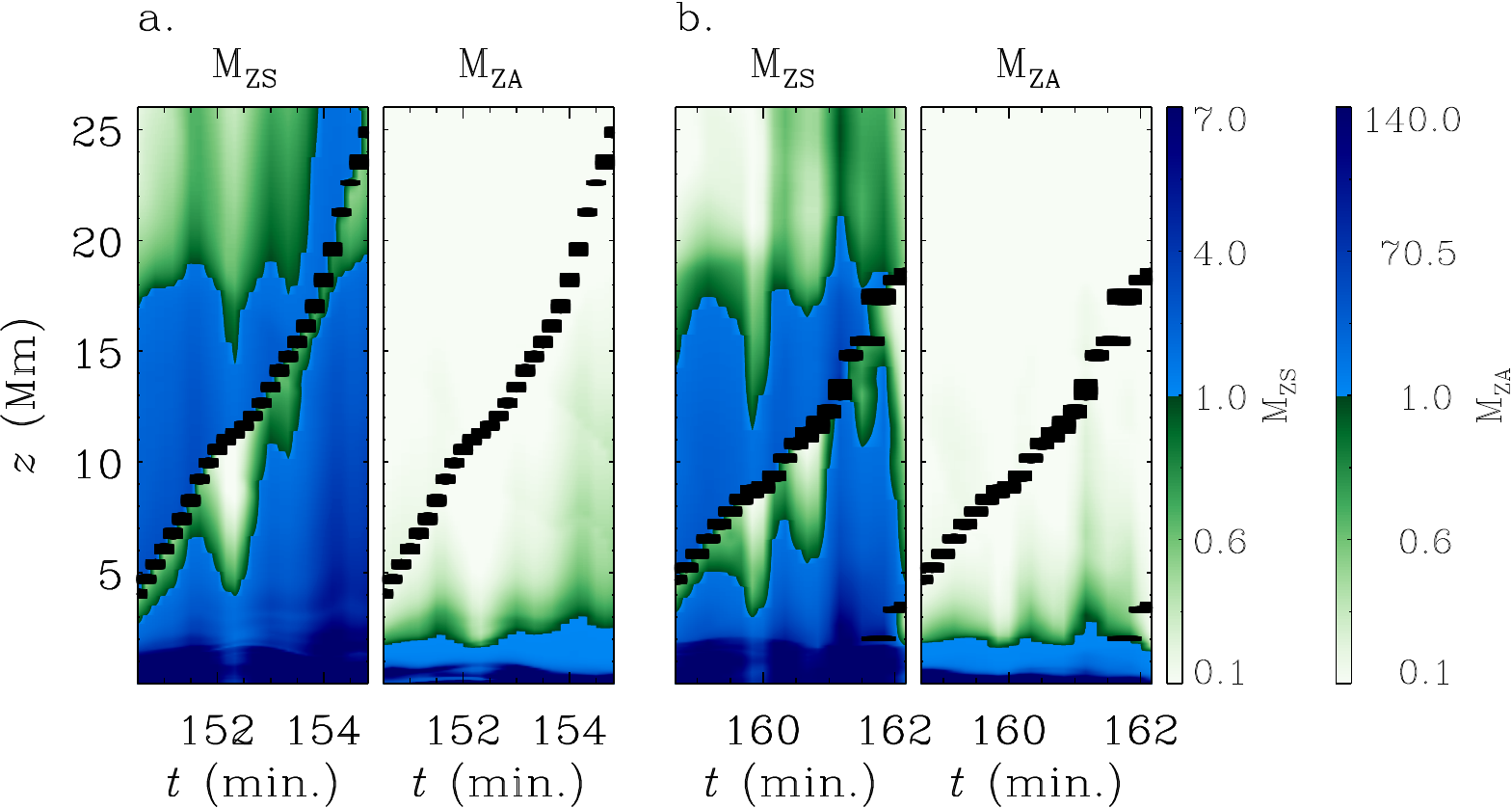} 
\caption{Time-distance plots (in the shock reference frame) of the $z$- sonic Mach number $ (M_{ZS} )$, and the $z$- Alfv\'en Mach number $ (M_{ZA}) $, along a vertical slit, for (a) $x$=-2.0\,Mm, and (b) $x$=8.2\,Mm. The shock fronts (of finite width) are overplotted using black rectangles. The flow speed transitions from values greater than the characteristic wave speed in the medium (slow wave speed, or sound speed for low-$\beta$) to values lower than it, inside the discontinuity. The discontinuity is evidently a slow MHD shock wave. 
}
\label{fig:mach_sframe}
\end{figure}

Further, we also measured the vertical speeds of propagation of these shock fronts.
For this purpose, we focused on a vertical slit at the fixed $x$-position where the tip of the shock front was located (see Table \ref{table:shock_slit}), and tracked the $z$-location of the shock front tip within that slit for each successive snapshot, starting from the one wherein the front was selected. The resulting $z-t$ curve for shock tip was then used to derive the shock propagation speed. A typical case is shown in Fig.~\ref{fig:shock_speed}. The resulting speeds lie in the range 25-190\,km\,s$^{-1}$. The measured shock speeds also allowed us to calculate the Mach numbers for the vertical flow in the reference frame of the shock, as follows. 
The vertical plasma flow velocity ($u_{z}$) was transformed into the shock frame ($SF$) from the lab frame ($LF$), using the relation: 
\begin{displaymath} u_{z,SF} =u_{z,LF} - u_{shock,LF}, \end{displaymath}

\noindent
where, $u_{shock, LF}$ is the speed of the shock in the lab frame.
Thereafter, we calculated the sonic Mach number, defined as
\begin{displaymath} M_{ZS} = \lvert u_{z,SF}\rvert/c_{s},  \end{displaymath} where $c_{s}$ is the local sound speed, and the Alfv\'en Mach number, defined as
\begin{displaymath} M_{ZA}= \lvert u_{z,SF}\rvert/v_{A}, \end{displaymath}
where $v_{A}$ is the local ($z$-) Alfv\'en velocity (i.e., normal to the shock front). The resulting time-distance plots of $ M_{ZS}$ and $ M_{ZA}$ along the vertical slit for a typical case are shown in Fig.~\ref{fig:mach_sframe}. 
It is known that for a slow MHD shock, under the assumption that \emph{slow MHD speed} $\sim$ \emph{sound speed} (low-$\beta$ regime), the following is satisfied \cite[e.g.,][]{LANDAU1984225}:
\begin{displaymath} M_{ZS,\mathrm{up}}> 1, M_{ZS,\mathrm{dn}} < 1, \end{displaymath}
\begin{displaymath}  M_{ZA,\mathrm{up}}< 1, \end{displaymath}
where $M_{ZS,\mathrm{up}}$ and $M_{ZS,\mathrm{dn}}$ denote the upstream and downstream sonic Mach numbers, respectively, while $M_{ZA,\mathrm{up}}$ denotes the upstream Alfv\'en Mach number (in the rest frame of the shock).
The above condition clearly holds good here (Fig.~\ref{fig:mach_sframe}). This further validates that the selected fronts are indeed slow MHD shock fronts, propagating along the direction of the local magnetic field.

\section{Results} 
\label{S-Results} 

\subsection{Nature of Shocks from Rankine-Hugoniot Relations} \label{S-RH cond}
We approximated the upstream sonic Mach numbers from the density jumps across the shock (for cases listed in Table \ref{table:shock_slit}), using the formula derived from Rankine-Hugoniot (RH) jump conditions for parallel MHD shocks:
\begin{equation} \label{eq:machno_jump}
r=\frac{(\gamma+1)M_{ZS,\mathrm{up}}^{2}}{(\gamma-1)M_{ZS,\mathrm{up}}^{2}+2},
\end{equation}
where $r=\frac{\rho_\mathrm{dn}}{\rho_\mathrm{up}}$ denotes the compression ratio (i.e., ratio of downstream density, $\rho_\mathrm{dn}$, to upstream density, $\rho_\mathrm{up}$, for any given shock). 
The relation \ref{eq:machno_jump} can be inverted to give:
\begin{equation} \label{eq:machno_jump_inverted}
M_{ZS,\mathrm{up}}^{2}=\frac{2 r}{r+1-\gamma(r-1)}.
\end{equation}
The values of $\rho_\mathrm{up}$ and $\rho_{\mathrm{dn}}$ were determined by the density at the tip and at the bottom of the shock (along the vertical slit), respectively, based on our shock detection criterion. 

The theoretically calculated upstream Mach numbers from Eq.~\ref{eq:machno_jump_inverted} for adiabatic ($\gamma=1.67$) and isothermal ($\gamma=1.0$) shocks were then compared with the values calculated directly (at the upstream point or the shock tip) using the measured speeds in the simulation (see Fig.~\ref{fig:jumps_RH}). Based on Fig.~\ref{fig:jumps_RH}, the adiabatic shock approximation is a valid one for describing the nature of the shocks in the simulation. It is to be noted, however, that the Rankine-Hugoniot conditions hold only for ideal cases and do not take into account effects of thermal conduction, radiation transfer or dissipation in the medium.

\subsection{Role of Shocks in Accelerating Plasma}

\begin{figure}[ht!] 
\includegraphics[width=0.5\textwidth]{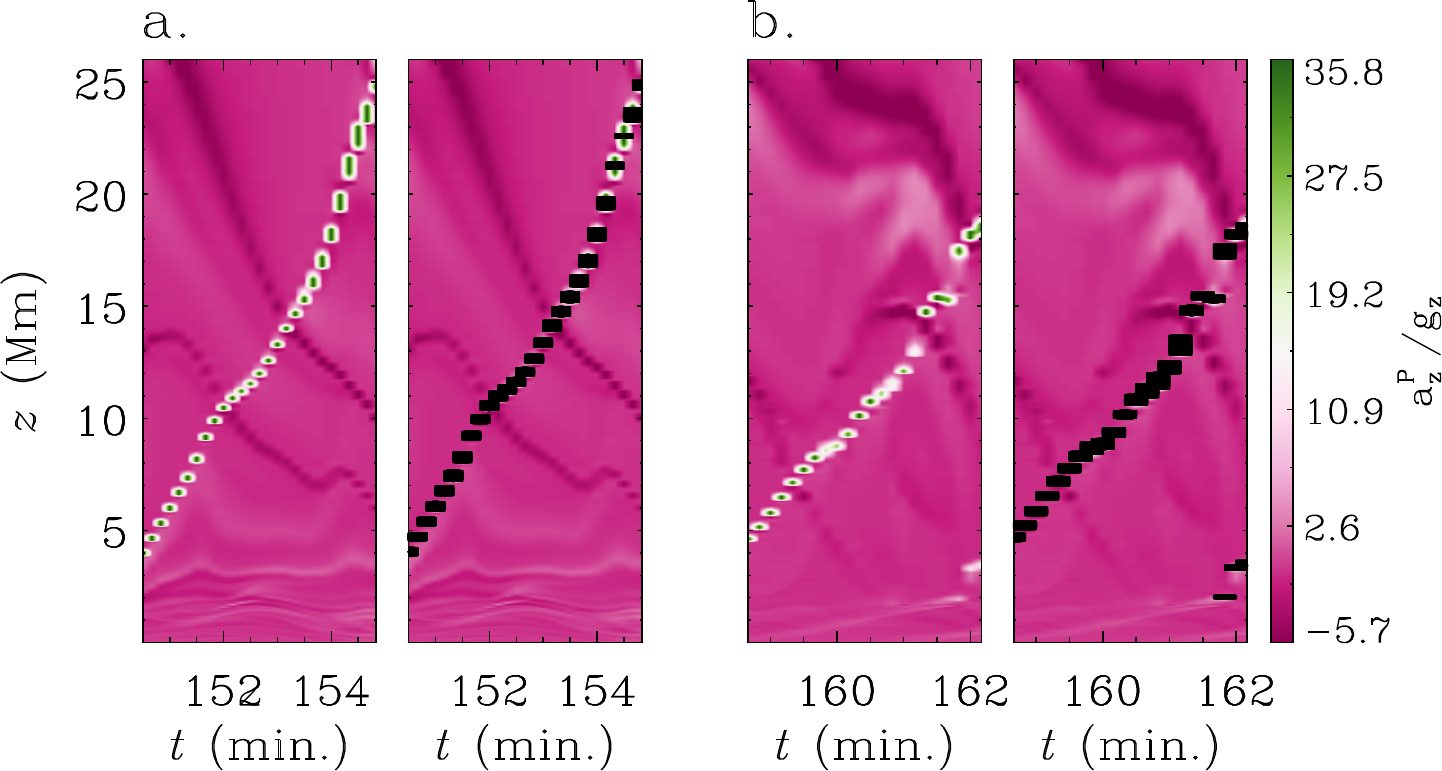} 
\caption{Vertical acceleration time-distance plots for two different horizontal positions, (a) $x$= -2.0\,Mm, and (b) $x$= 8.2\,Mm. The shock fronts are overplotted in black. Notice that the shock fronts coincide with regions of strong positive vertical acceleration of the plasma.}
\label{fig:acc_timedist}
\end{figure}

For each of the selected shock fronts, let us again focus on a vertical slit at the fixed $x$-position, where the tip of the shock front was located, and this time we constructed a time-distance plot of the vertical ($z$-) component of plasma acceleration along the slit (see Fig.~\ref{fig:acc_timedist}). It is to be noted that the $z$-acceleration considered here includes only the first two terms of equation (\ref{eq:NS}), i.e., $a_z^P$ given by,
\begin{equation} \label{eq:m}
a_{z}^P = \frac{- \partial p/ \partial z} {\rho} - g_{z} 
\end{equation}

Next, we overplotted the shock fronts on these time-distance plots. From these plots, we can see that the shock fronts coincide with regions of strong positive vertical acceleration of the plasma (caused by strong pressure gradient across the shock). It is the same plasma that forms the tip of the spicule material during its rise phase, as the time-distance plots for synthetic intensity indicate (see Fig.~\ref{fig:Intensity_timedist}). This suggests that the shock fronts are energizing the plasma and aiding its upward rise in the form of a jet-like structure. Note that when we make the acceleration time-distance plots with the additional acceleration term due to Lorentz force included, we do not see a qualitative difference in the appearance of the plots, that is, the shock fronts still coincide with regions of strong positive vertical acceleration.

\subsection{Finding Maximum Height Reached by Spicules}

\begin{figure}[ht!] 
\includegraphics[width=0.5\textwidth]{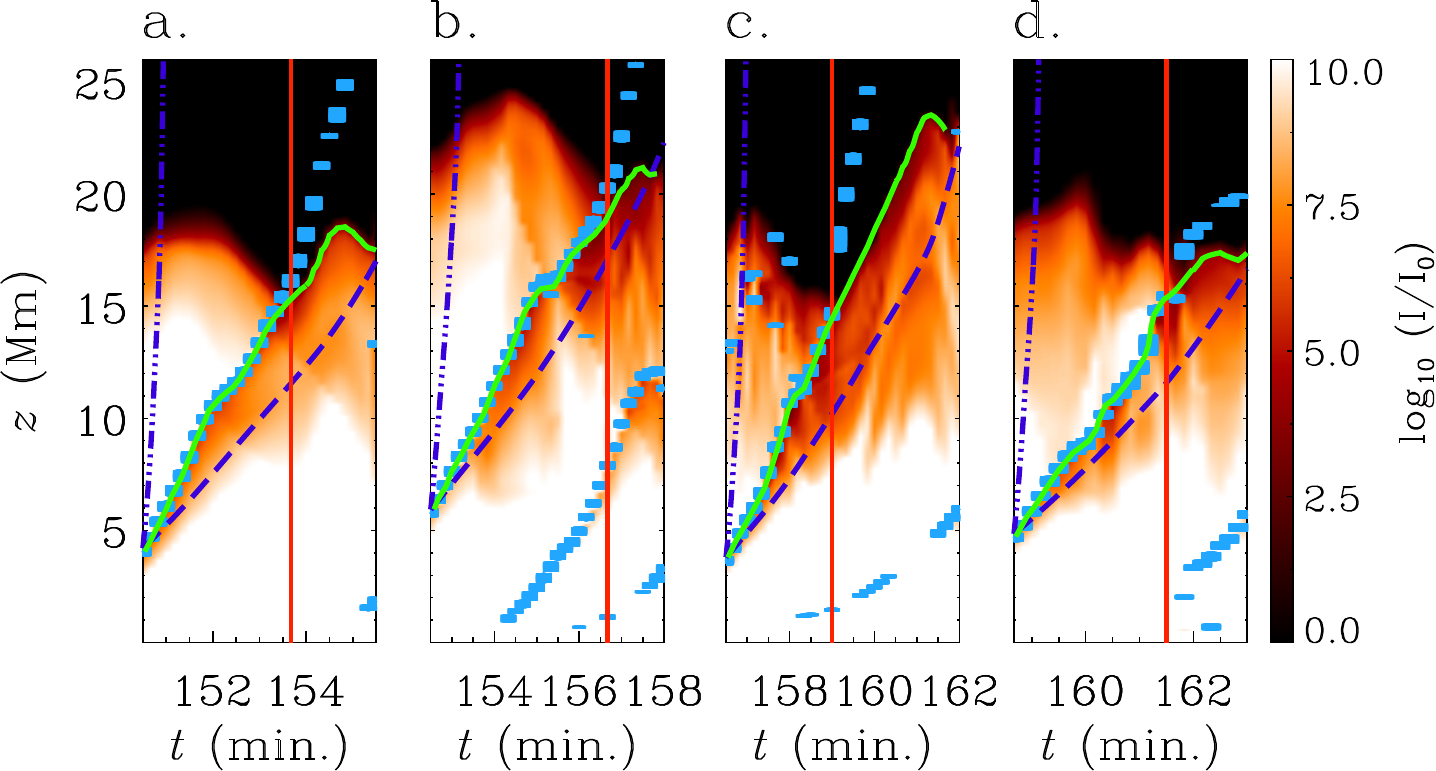} 
\caption{Synthetic intensity (for emission at $T_{\lambda} = 15,000 \text{K}$) time-distance plots for different horizontal positions, a. $x$= -2.0\,Mm, b. $x$= 2.2\,Mm, c. $x$= -4.1\,Mm, and d.  $x$= 8.2\,Mm. The shock fronts are overplotted in light blue. The dark blue dashed line shows the sound speed characteristic, and the dark blue dot-dashed line is for the Alfv\'en speed characteristic. The vertical line is at the time instant after which the shock front and spicule trajectory separate. The spicule trajectory is shown in green.}
\label{fig:Intensity_timedist}
\end{figure}

For each of the selected shock fronts, the maximum height reached by corresponding spicule was also determined using a time-distance plot of the synthetic intensity (for emission at $T_{\lambda} = 15,000\,\text{K}$) constructed along 
the same vertical slit as before (see Table \ref{table:shock_slit}). The time-distance plots clearly show the corresponding 
spicule (as seen through its synthetic emission) rising and falling with time, along that 
slit (see Fig.~\ref{fig:Intensity_timedist}).  On each time-distance plot, 
we traced out the trajectory of the spicule based on visual inspection (shown in green). 
The maximum height reached by the spicule above the photosphere was estimated by 
finding the apex of this trajectory in $t-z$ space. 
\begin{table}[h]
\centering
\begin{tabular}{clcclclc} 
 \hline
 \hline
 \multirow{2}{1em} {No.}
 &
 \multirow{3}{2em}{$t$ (min.)} & \multirow{3}{2em} {$x_\mathrm{loc}$ (Mm)} & \multirow{2}{2em}{ $h_{max}$ (Mm)} & \multirow{3}{2em} {$\frac{a_{z}^P}{g_{z}}$} & \multirow{3}{2em} {$\frac{a_z^L}{g_z}$} &  \multirow{3}{1.5em} {$r=\frac{\rho_\mathrm{dn}}{\rho_\mathrm{up}}$} & \multirow{3}{3em} {$M_{ZS,\mathrm{up}}$ (meas.)}\\ 
 &  &  &  &  & & &\\
&  &  &  &  & & &\\
\decimals
 \hline
1. & 146.6 & 3.3 & 15.3 & 33.9 & 1.1 & 2.6 & 2.4\\ 
2. & 148.0 & 8.8 & 19.1 & 65.5 & -21.0 & 3.1 & 3.0\\ 
3. & 150.2 & -4.2 & 19.6 & 36.9 & -14.6 & 2.7 & 2.8\\
4. & 150.5 & -2.0 & 18.5 & 41.2 & -9.6 & 2.7 & 2.6\\ 
5. & 152.5 & 2.2 & 21.2 & 58.3 & -2.6 & 2.2 & 2.1\\ 
6. & 154.7 & 3.8 & 17.5 & 20.5 & 4.9 & 2.4 & 1.9\\
7. & 154.8 & 7.6 & 6.9 & 15.0 & -3.1 & 1.9 & 2.0\\
8. & 155.3 & 2.2 & 11.7 & 22.1 & 0.9 & 2.4 & 1.8\\
9. & 156.2 & -1.1 & 24.9 & 46.7 & -18.5 & 3.3 & 3.2\\ 
10. & 156.5 & -4.1 & 23.5 & 28.9 & 0.4 & 2.7 & 2.2\\
11. & 156.8 & 1.3 & 8.6 & 15.9 & 0.8 & 1.5 & 1.9\\
12. & 157.5 & 0.8 & 22.3 & 35.6 & -16.7 & 2.6 & 2.4\\
13. & 158.7 & 8.2 & 17.4 & 30.0 & -6.7 & 2.6 & 2.1\\
14. & 160.5 & -1.1 & 6.3 & 21.4 & -2.0 & 2.2 & 2.1\\
 \hline
\end{tabular}
\caption{Properties of the 14 selected shocks and the corresponding spicules. The time, $t$, of the snapshot wherein the shock front was selected is shown, along with the $x-$ location of slit ($x_\mathrm{loc}$). The maximum height of spicule is labelled as $h_{max}$. The correlation coefficient (Spearman) for $h_{max}$ vs.~$a_{z}^P/g_{z}$ (inside the shock) = 0.70 (p-value: 0.006), while the correlation coefficient (Spearman) for $h_{max}$ vs.~$a_{z}^L/g_{z}$ (inside the shock) = -0.50 (p-value: 0.069). The upstream sonic Mach numbers calculated from measurements, labelled as $M_{ZS,\mathrm{up}}$ (meas.), are all $>1$, as expected.}
\label{table:shock_slit}
\end{table}

\begin{figure}[h]
\centering
\includegraphics[width=0.45\textwidth]{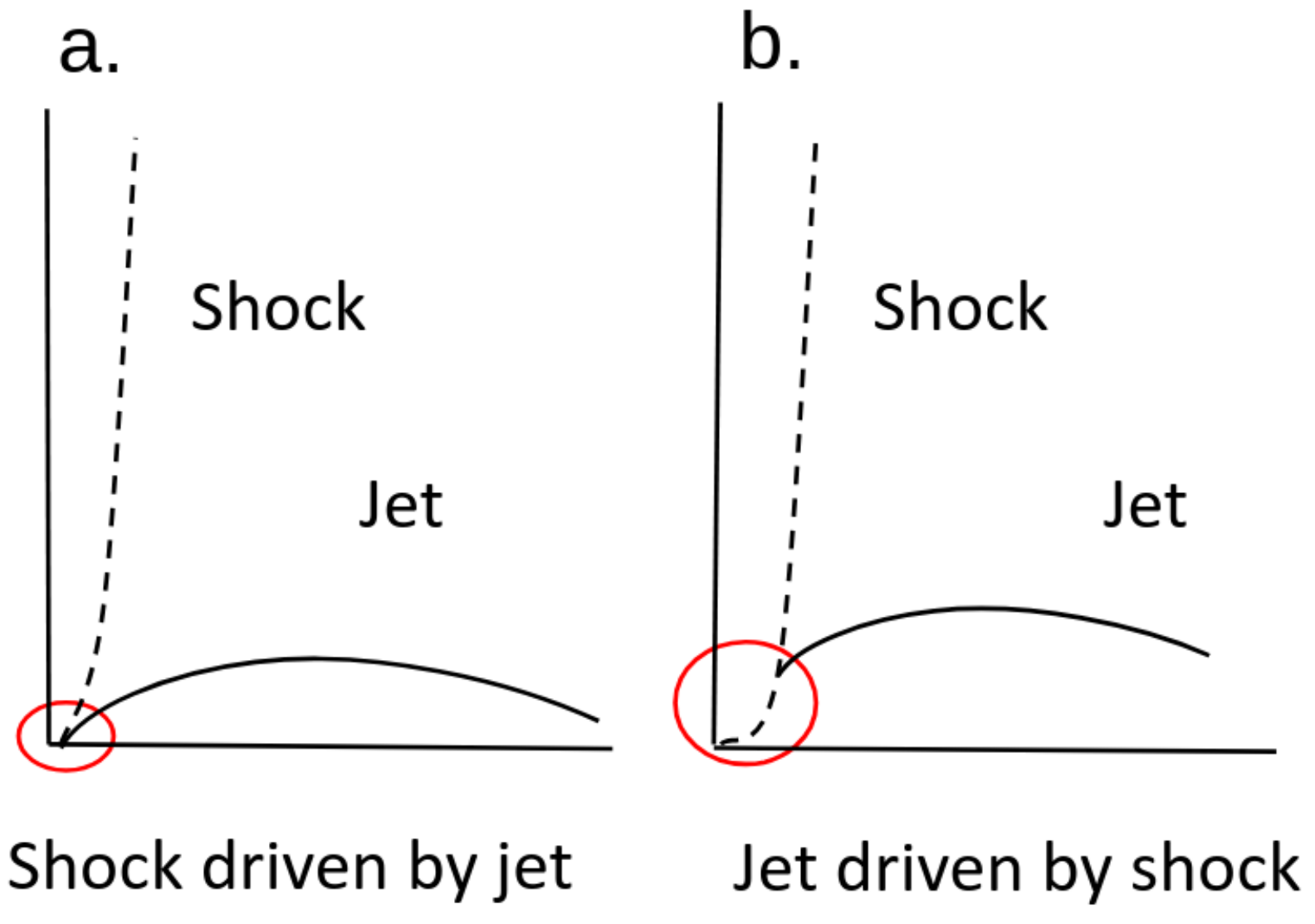} 
\caption{Illustration of the difference in time-distance curves of (a) a shock driven by a jet, and (b) a jet driven by shock. The red ellipses are to guide the eye towards the dissimilarity that the trajectories of shock and jet coincide only at initial time for the case (a), while they coincide for an extended time for the case (b).}
\label{fig:shock_jet_cartoon}
\end{figure}

We also overplot the shock fronts, the sound speed and the Alfv\'en speed characteristic curves in the same Fig.~\ref{fig:Intensity_timedist}. 
It may be further noted here that the shock moves together with the rising plasma material 
for some time before its trajectory separates from the parabolic trajectory of the spicule material (see panel (b) of Fig.~\ref{fig:shock_jet_cartoon}). This scenario is different from a shock wave that is created by a 
high-speed supersonic jet shocking the overlying medium. Such a shock trajectory would quickly move upwards away from the jet trajectory once created and not show any overlap (see panel (a) of Fig.~\ref{fig:shock_jet_cartoon}). Also, the jump for a shock that is created by a supersonic jet will propagate ahead of the jet, rather than encompassing the jet tip inside the discontinuity as in our case.

\subsection{Characterizing the Effect of Shocks} \label{S-Characterizing shock effects}

In order to characterize the relationship between shocks and spicules further in quantitative terms, we calculated the mean vertical plasma acceleration inside the shock front regions within the slit, at the snapshot just before the shock front and spicule separate (see vertical lines in Fig.~\ref{fig:Intensity_timedist}). Then, we correlate this parameter with the overall height reached by the spicule.

\begin{figure}[ht!] 
\includegraphics[width=0.48\textwidth]{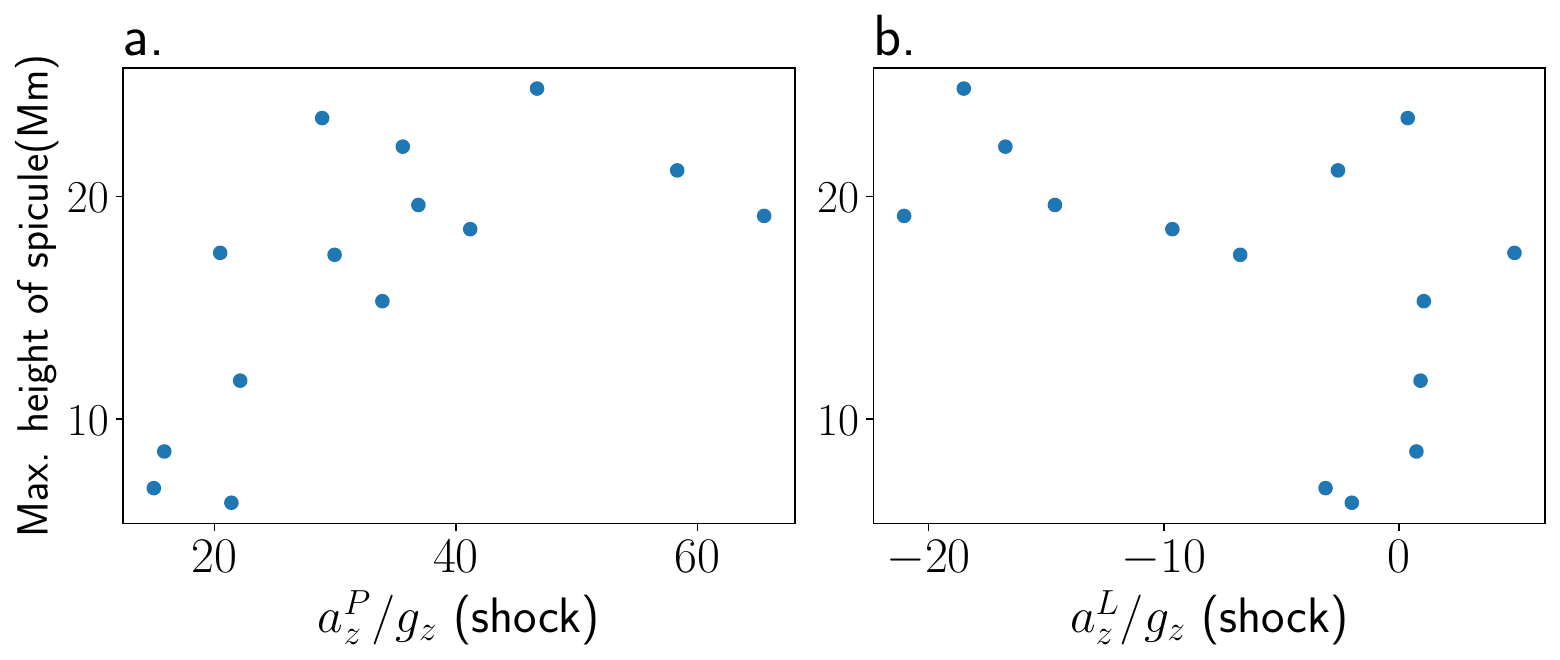} 
\caption{(a) Maximum height of spicule vs.~$a_{z}^P/g_{z}$ for the shock front. Notice the positive correlation (correlation coefficient=0.70) between the two quantities. (b) Maximum height of spicule vs.~$a_{z}^L/g_{z}$ for the shock. Notice the negative correlation (correlation coefficient=-0.50) between the two quantities.}
\label{fig:azshock_corr}
\end{figure}

The calculated mean $a_{z}^P/g_{z}$ values inside the shock show a positive correlation with the maximum height reached by the corresponding spicule. This shows that the strength of shocks may play a vital role in determining the heights of the spicules, supporting the idea that MHD shocks act as drivers of spicules (see Fig.~\ref{fig:azshock_corr} for cases in Table \ref{table:shock_slit}).  In other words, a stronger shock, which energizes the plasma more, tends to drive a taller spicule. 

On the other hand, if we compute the mean value of vertical acceleration inside the shock front due to Lorentz force (third term of Eq.~\ref{eq:NS}) scaled by solar gravity, $(\JJ\times\Bv)_{z}/\rho g_{z} (\equiv a_{z}^L/g_{z})$, at the time of separation of the shock and the spicule trajectories, and then correlate this with the spicule height, we obtain a weak negative correlation (see Fig.~\ref{fig:azshock_corr} for cases in Table \ref{table:shock_slit}). This means that it is primarily the strong pressure gradient across the shock that is responsible for lifting the plasma material in the form of a spicule jet (since gravity only contributes towards decelerating the jet). The above conclusion holds for the spicules driven by acoustic shocks formed in our model either due to leaking p-modes or convection. Note that there may be other classes of spicules e.g., generated by magnetic reconnection.

\subsection{Propagating Disturbances in the Corona}

\begin{figure}[h]
\centering
\includegraphics[width=0.4\textwidth]{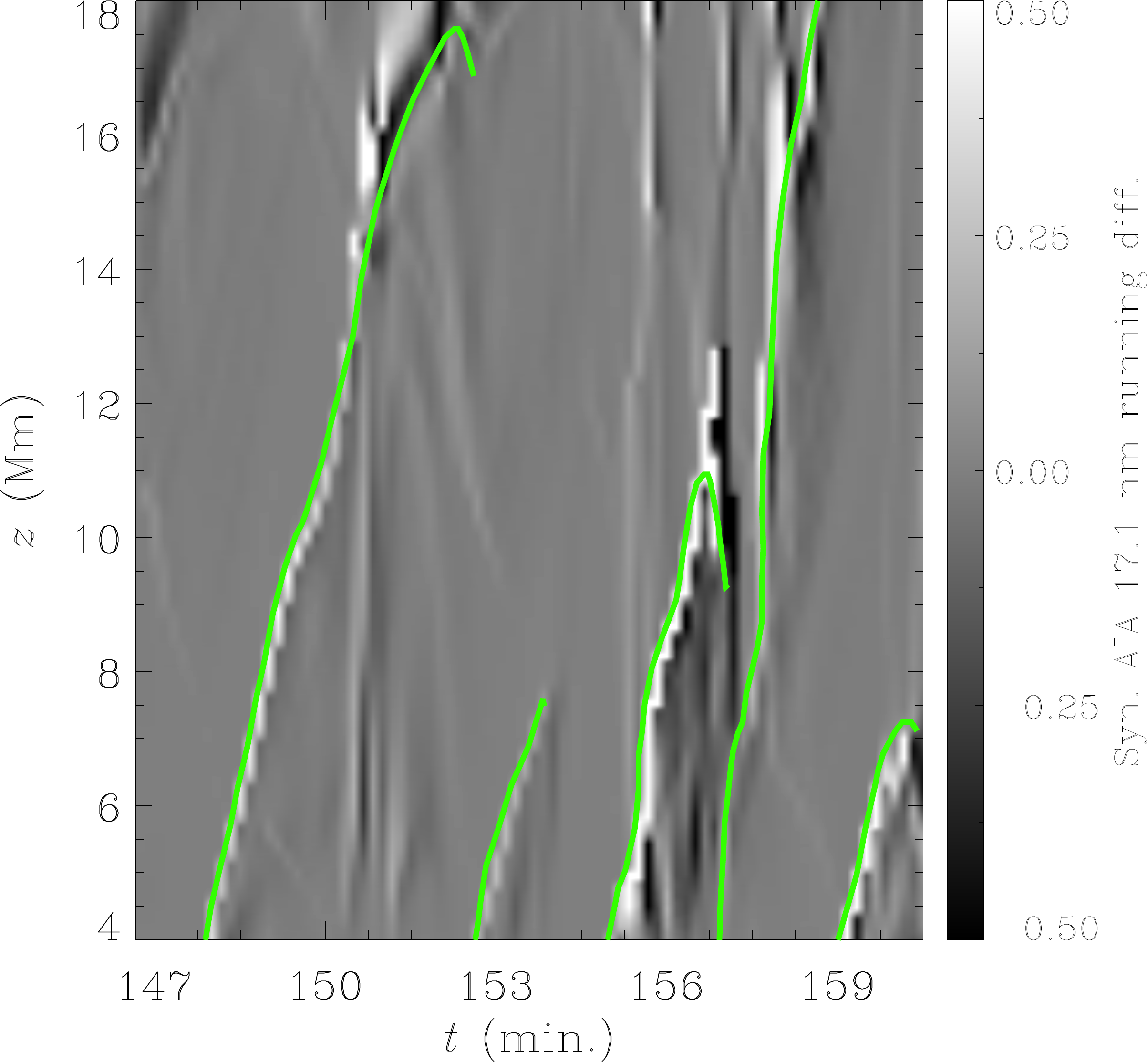} 
\caption{Time-distance plot of synthetic AIA 17.1 nm intensity running difference along a slit at $x$= 8.8\,Mm. Notice the intensity enhancements propagating upwards. The rising phase of the spicule trajectory, as seen through the synthetic emission at $T_{\lambda} = 15,000 \text{K}$, is shown in green.}
\label{fig:aia171_pcd}
\end{figure}

We synthesized the emission in the SDO/AIA 17.1 nm line using the Eq.~\ref{eq:syn}, with the contribution function $G_{\lambda}$ obtained from the SolarSoft library (and interpolated for the temperature values for the snapshot). The SolarSoft library is available at \url{https://www.lmsal.com/solarsoft/}. The synthetic emission was scaled by its horizontal average for each height in a given snapshot. Then, we constructed a time-distance plot of running-difference of AIA 17.1 nm emission along the same vertical slit as before. A resulting plot is shown in Fig.~\ref{fig:aia171_pcd}. Here, we see the presence of structures similar to PCDs \citep{jiao2015ApJ...809L..17J,samanta2015ApJ...815L..16S,bryans2016ApJ...829L..18B,DePontieu2017ApJL,bose2023ApJ...944..171B,Skirvin2024A&A...689A.135S}, which are
associated with shock waves driving spicules (coinciding temporally with the rising phase of the spicules) that subsequently propagate into the corona. It would be interesting to further study the properties of these intensity (density) enhancements seen in the simulation and reported in the observations.

\section{Discussion and Conclusions} 
      \label{S-Discussion}   
      
In this paper, we analyzed the data from a high-resolution 2D MHD simulation model of the stratified solar atmosphere spanning from the upper convection zone to the lower corona, where spicules are self-consistently excited due to the subsurface convective processes. 

The model includes LTE radiative transfer, an ionized equation of state with ionization calculated using Saha ionization formula, and several important physical processes such as anisotropic thermal conduction along field lines, semi-relativistic Boris correction to Lorentz force, Ohmic and viscous heating, turbulent diffusion and optically thin radiative cooling. Slow MHD shocks are generated in the simulation by non-linear wave steepening (due to sharp density decrease) as the slow modes propagate upward through the chromosphere and transition region. The modes may themselves be excited due to a variety of mechanisms like granular squeezing and solar global modes \cite[for a discussion, see][]{dey2022polymeric}. We identified the shocks directly and selected a total of 14 cases as random but typical samples for further investigation. 

The shocks and spicules follow the magnetic field lines. By tracking the shocks and the corresponding spicules along a vertical slit, we found 
that the shocks coincide spatio-temporally with the tip of the spicule during its rise phase. 
In all the cases simulated, we observe that the shock drives the spicule. We could then also quantify the effects of the various forces on the acceleration of the plasma 
inside the spicules during their rise. 
It was found that primarily the pressure gradient across the shock has a dominant contribution in lifting the plasma material in the form of a 
spicular jet, as seen in the time-distance plots for vertical acceleration in Fig.~\ref{fig:acc_timedist}. 

The shock fronts, as they propagate upward, energize the local plasma to rise along 
with them in the form of a jet \cite[see also Fig.~3 and Extended Data Fig.~7 in][]{dey2022polymeric}. There is a positive correlation between the mean vertical acceleration of the plasma inside the shock due to the pressure gradient (before they separate), and the height of the spicule (in the range 6-25 Mm). A stronger shock, which energizes the plasma more, tends to drive a taller spicule. 

Further, we identified that the mean vertical acceleration due to Lorentz force inside the shock fronts similarly has a weak negative correlation with spicule height. This might also be related to the finding of shorter overall heights of jets in regions of strong magnetic fields as found in simulation by \cite{Kesri2024ApJ...973...49K}. However, we acknowledge that the reported correlations are 
limited by the small size of our sample. Another factor that may influence the result is the selection of a vertical slit of width one grid spacing, which ignores the transverse motion of the spicules and shocks. The measurement of heights may be sensitive to the slit width. Nevertheless, the acceleration time-distance plots, as well as the positive $a_{z}^P/g_{z}$ and (mostly) negative values $a_{z}^L/g_{z}$, inside the shock, still validate our interpretation of the effects of the various forces on lifting the plasma material. The conclusion on shock driving of the spicule will also remain unaffected. The slow MHD shocks driving the spicules move together with these jets up into the lower coronal heights. Subsequently, the shocks separate and continue to travel further upwards 
into the corona as density perturbations while 
the jet falls back due to gravity. 
We could also see the corresponding intensity enhancements in the synthesized AIA 17.1 nm emission. 
This behaviour seems to be similar to that of the PCDs. The analogy is further strengthened by the fact that the shock speeds as found in this study ($\sim$ 100--150\,km\,s$^{-1}$) are similar to those reported for PCDs, e.g., \cite{samanta2015ApJ...815L..16S}. We will carry out a detailed investigation in future to study this further.

In conclusion, this study using data from a model of solar spicules that is self consistently driven by solar convection leading to excitation of slow MHD shocks, shows that:
\begin{enumerate}
\item Slow MHD shocks in the low-beta solar plasma atmosphere are regions of strong positive vertical acceleration of the plasma that forms the tip of the spicule material during its rise phase, and later escape into the low corona;
\item Heights of the jets, at least in open magnetic field regions, are determined by the strength of shocks driving them;
\item Primarily, strong pressure gradient across the shock helps in lifting the plasma material, and determines the heights of the spicules; 
\item In our model with the imposed vertical magnetic field of 74\,G, the strength of the Lorentz force inside the shock front appears to plays a subdominant role in determining the spicule heights, even though the magnetic field in the corona determines the direction of the jetting.
\end{enumerate}

We would like to 
remark that while the exact 
values of spicule heights 
and strength of the 
corresponding shocks will 
be dependent on the 
properties of the dynamical 
atmosphere through which 
the shocks propagate, our 
view is that the 
correlation between the two 
in a given model is a 
robust result. In the 
future, we plan to study 
different models by varying 
the cooling function 
\cite[e.g., by using 
recipes outlined in][] 
{carlsson2012A&A} and by 
modeling the heating due to 
magnetic field turbulence 
to study, for example, the 
effect of these on the 
thickness of the TR and the 
strength of shocks. It has 
been shown recently that 
the thickness of TR is 
dependent on the non-equilibrium ionization 
effects that require a very 
high grid resolution 
\citep{Matsumoto2025ApJ}.

Finally, it is only pertinent to raise the following questions: i) What would be the fate of these shocks as they continue their (apparently) unimpeded propagation into the corona? What would be their contribution to (ii) the temperature and momentum distribution in the low solar corona, or iii) to the solar wind? It remains to be seen, both theoretically and observationally, how these important questions will be settled in the future.

\section{Acknowledgments}

\noindent

We thank the referee for careful reading of our paper and providing comments that have considerably improved the quality of the presentation. We also thank Dr Tanmoy Samanta for useful discussions and insights. Computing time provided by Nova HPC at IIA is gratefully acknowledged. Further, this work used the DiRAC Data Intensive service (DIaL2) at the University of Leicester (project id: dp261), managed by the University of Leicester Research Computing Service on behalf of the STFC DiRAC HPC Facility (\href{www.dirac.ac.uk}{www.dirac.ac.uk}). The DiRAC service at Leicester was funded by BEIS, UKRI and STFC capital funding and STFC operations grants. DiRAC is part of the UKRI Digital Research Infrastructure. This research also made use of SciPy \citep{Virtanen_2020}, Matplotlib \citep{Hunter:2007} and NumPy \citep{harris2020array} libraries in Python. R.E. acknowledges the NKFIH (OTKA, grant No. K142987) Hungary for enabling this research. 
R.E. is also grateful to Science and Technology Facilities Council (STFC, grant No. ST/M000826/1) UK, PIFI (China, grant No. 2024PVA0043) and the NKFIH Excellence Grant TKP2021-NKTA-64 (Hungary).
This work was also supported by the International Space Science Institute project (ISSI-BJ ID 24-604) on "Small-scale eruptions in the Sun". 

\section{Code Availability}

The Pencil Code can be downloaded from \dataset[pencil-code.org]{https://pencil-code.org}. The MHD setup used to generate the reported simulation data is publicly available at: 
\dataset[doi:10.5281/zenodo.5807020]{https://doi.org/10.5281/zenodo.5807020}. 
\bibliography{shocks_main}{}
\bibliographystyle{aasjournal}

\end{document}